\documentclass[
]{ceurart}

\usepackage{listings}
\usepackage{mdframed}
\usepackage{enumitem}
\usepackage{graphicx}
\usepackage{multirow}
\usepackage{todonotes}
\usepackage{hyperref}
\definecolor{link}{rgb}{0.63, 0.79, 0.95}

\begin{document}

\copyrightyear{2026}
\copyrightclause{Copyright for this paper by its authors. Use permitted under Creative Commons License Attribution 4.0 International (CC BY 4.0).}

\conference{Joint Proceedings of REFSQ-2026 Workshops, Doctoral Symposium, Posters \& Tools Track, and Education and Training Track. Co-located with REFSQ 2026. Poznan, Poland, March 23-26, 2026}

\title{The Ethos of the PEERfect REVIEWer:\\ Scientific Care and Collegial Welfare}
\subtitle{With an Accompanying Guideline for Peer Review with Scientific Rigor and Empathy}


\author[1]{Oliver Karras}[%
orcid=0000-0001-5336-6899,
email=oliver.karras@tib.eu,
]
\cormark[1]

\address[1]{TIB - Leibniz Information Centre for Science and Technology, Germany}

\cortext[1]{Corresponding author.}

\begin{abstract}
\textit{\textbf{[Background.]}}
Peer review remains a cornerstone in academia, including software engineering and requirements engineering, yet it frequently falls short in fostering joint progress and well-being. While peer review primarily emphasizes scientific rigor, it often lacks the empathy essential for supporting and encouraging all peers involved --- risking emotional stress, discouragement, or even withdrawal, particularly among junior and early-career researchers.
\textbf{[Objective.]}
In this experience report, I aim to highlight that peer review is a practice that demands both scientific care for quality and collegial welfare for the joint progress and well-being of all peers involved, including authors, co-reviewers, workshop or conference organizers, and journal editors.
\textbf{[Method.]}
Drawing on my ten years of experience in academia, including over 100 publications, over 175 self-authored reviews, and over 25 organizational roles, I propose the ethos of the PEERfect REVIEWer, grounded in the two core values: \textit{Scientific care} and \textit{collegial welfare}. Through reflection shaped by professional exchanges with colleagues, consideration of literature, and an examination of both self-authored and received reviews, I formulated an accompanying guideline with 16 practical recommendations to guide reviewers in their actions to achieve these two values.
\textbf{[Results.]}
The ethos of the PEERfect REVIEWer and its accompanying guideline help reviewers in upholding high scientific standards and conducting peer review in a constructive, supportive, respectful, and timely manner. They demonstrate that scientific rigor and empathy are complementary forces that promote impactful peer review practice.
\textbf{[Conclusions.]}
By placing scientific care and collegial welfare at the core of peer review, this experience report reaffirms the importance of scientific rigor while also advocating for greater attention to empathy. It invites reviewers to reconsider their role not merely as gatekeepers but as partners in the academic journey of each peer involved. The PEERfect REVIEWer is both a caretaker of quality and a steward of joint progress and well-being --- as truly impactful peer review practice requires scientific rigor and empathy in equal measure.
\end{abstract}




\maketitle

\section{Introduction}
Peer review is a cornerstone in academia~\cite{Mulligan.2013, Prechelt.2018, Shah.2022, Sridhar.2025}. Despite its relevance, peer review is often prone to shortcomings such as opaque decision-making, inconsistent standards, and discouraging feedback~\cite{Shah.2022, Ozkaya.2020, Soldani.2020}. Reviews are frequently perceived as short, rushed, rude, unconstructive, overly subjective or even biased, and lacking expertise, explanations, and justifications~\cite{Prechelt.2018, Shah.2022, Ozkaya.2020, Soldani.2020, Ernst.2021}.

Given the growing awareness of these issues, several researchers investigated peer review in software engineering (SE)~\cite{Prechelt.2018, Shah.2022, Ozkaya.2020, Soldani.2020, Ernst.2021}. These works consistently emphasize scientific care for quality as the central value of peer review, while collegial welfare for joint progress and well-being of all peers involved, including authors, co-reviewers, workshop or conference organizers, and journal editors, is treated as a secondary concern. Peer review, including its organizational efforts, is a voluntary service to the community. Yet, it contributes little to professional reputation~\cite{Prechelt.2018, Shah.2022, Ozkaya.2020, Soldani.2020}. In addition, the number of submissions is constantly increasing, which also increases the workloads of all peers involved~\cite{Prechelt.2018, Ozkaya.2020, Soldani.2020, Ernst.2021}. Prechelt et al.~\cite[p. 78]{Prechelt.2018} also found that reviewers perceive ``\textit{help}[ing] \textit{authors to become better researchers}'' as one of the least important main purposes of peer review. Similarly, Ernst et al.~\cite{Ernst.2021} found that reviewers perceive \textit{kindness} as one of the least important characteristics of good reviews. All these findings substantiate the predominantly technical consideration of peer review as a gatekeeping mechanism for ensuring scientific rigor~\cite{Prechelt.2018, Sridhar.2025, Ernst.2021}. While scientific care for quality is essential, this narrow framing overlooks the fact that peer review is, at its core, a human endeavor --- one that requires collegial welfare for joint progress and well-being of all peers involved. When this value is neglected, the consequences are tangible: Unfair assessment, emotional stress, gradual discouragement, declining motivation, or even withdrawal, particularly affecting junior and early-career researchers~\cite{Sridhar.2025, Kismihok.2024}. These consequences are rarely intentional, but rather the result from an imbalance between scientific care and collegial welfare.

Other disciplines, e.g., marketing~\cite{Sridhar.2025, Houston.2021} or medicine~\cite{Gregory.2019, Mavrogenis.2020}, have made notable progress by reframing peer review as a collaborative opportunity in which all peers involved work toward accepting publications rather than searching for reasons to reject them. However, SE, including requirements engineering (RE), is lagging behind. Although RE is part of SE and shares its peer review culture, expectations, and challenges, to the best of my knowledge no peer-reviewed work has addressed peer review within the RE community. According to Ernst et al.~\cite[p. 25]{Ernst.2021}: ``\textit{The grey literature --- blogs, commentaries in journals, editor’s notes, and so on --- are, in our view, still the best current source for explicit guidelines on how to do peer review.}'', which even emphasizes the general lack of peer-reviewed literature in SE.

For this reason, Ernst et al.~\cite{Ernst.2021} developed a proto-guideline with criteria for writing reviews based on interviews and an online survey of experienced reviewers. ``\textit{This guideline is for general characteristics of well-written peer reviews in software engineering}''~\cite[p. 21]{Ernst.2021} and serves as a starting point that ``\textit{is naturally far from being perfect, let alone complete}''~\cite[p. 27]{Ernst.2021}. This experience report builds on this preliminary guideline by integrating a human-centered perspective and extending its scope beyond review writing to encompass a broader peer review practice --- covering the themes \textit{Commitment \& Planning}, \textit{Review Preparation \& Execution}, \textit{Feedback \& Tone}, and \textit{Ethical Responsibility \& Boundaries}. In this way, it addresses the frequently stated need for explicit, experience-based guidance to improve peer review practice~\cite{Prechelt.2018, Ozkaya.2020, Ernst.2021, Shah.2022} and initiates this discussion within the RE community.

Drawing on a decade of experience in academia --- including over \href{https://www.oliver-karras.de/publications/}{\textit{\textcolor{link}{100 publications}}}, over \href{https://www.webofscience.com/wos/author/record/656994}{\textit{\textcolor{link}{175 self-authored reviews}}} for workshops, conferences, journals, and funding programs with \href{https://www.oliver-karras.de/about/}{\textit{\textcolor{link}{9 distinguished reviewer awards}}}, and over \href{https://www.oliver-karras.de/community-service-oc/}{\textit{\textcolor{link}{25 organizational roles}}} --- I propose the ethos of the PEERfect REVIEWer, grounded in the two core values: \textit{Scientific care} and \textit{collegial welfare}. Through reflection shaped by professional exchanges with colleagues, consideration of literature, and an examination of both self-authored and received reviews, I extended and reformulated the criteria by Ernst et al.~\cite{Ernst.2021} into an accompanying guideline with 16 practical recommendations to guide reviewers in their actions to achieve these two values. In this way, I aim to cultivate a new generation of PEERfect REVIEWers --- caretakers of quality and stewards of joint progress and well-being --- who promote impactful peer review practice for joint academic, professional, and personal progress through a constructive, ethically responsible, and empathetic peer review culture that advances our community.\vspace{0.2cm}


\section{Related Work}
\label{sec:rw}
Prior research on peer review in SE has extensively examined its technical dimensions, including characteristics and root causes of good and poor reviews~\cite{Prechelt.2018, Shah.2022, Ozkaya.2020, Soldani.2020, Ernst.2021}, effects of different peer review models~\cite{Prechelt.2018, Shah.2022, Ernst.2021}, review workload~\cite{Prechelt.2018, Ozkaya.2020, Soldani.2020, Ernst.2021}, and incentive mechanisms~\cite{Prechelt.2018, Shah.2022, Ozkaya.2020, Soldani.2020}. These works have been instrumental in identifying challenges such as inconsistent standards, reviewer fatigue, and lack of recognition, and have proposed solutions ranging from computational tools to preliminary guidelines. Although they also acknowledge the human costs of peer review — including emotional stress, gradual discouragement, declining motivation, or even withdrawal — they still predominantly frame peer review as a gatekeeping mechanism for ensuring scientific rigor and quality. As a result, empathy, joint progress, and well-being of all peers involved receive little attention. Even community-driven efforts to define good peer review practice tend to prioritize factuality and specificity over kindness and support~\cite{Ernst.2021}.

Beyond SE research, several professional organizations provide ethical or procedural guidance for reviewers. The \textit{COPE Ethical Guidelines for Peer Reviewers}~\cite{Cope.2017} focus on integrity, confidentiality, fairness, and responsible conduct, while publisher guidelines (e.g., \href{https://www.elsevier.com/reviewer/how-to-review}{\textit{\textcolor{link}{Elsevier}}}, \href{https://www.springernature.com/de/authors/campaigns/how-to-peer-review-2}{\textit{\textcolor{link}{Springer Nature}}}, \href{https://reviewers.acm.org/training-course/acm-peer-review-training-and-certification}{\textit{\textcolor{link}{ACM}}}, or \href{https://journals.ieeeauthorcenter.ieee.org/submit-your-article-for-peer-review/become-an-ieee-reviewer/}{\textit{\textcolor{link}{IEEE}}}) primarily outline procedural expectations and ethical responsibilities. Similarly, the \href{https://www2.sigsoft.org/EmpiricalStandards/}{\textit{\textcolor{link}{ACM Empirical Standards}}}~\cite{Ralph.2021} specify criteria for evaluating empirical research but do not offer practical recommendations on how to conduct peer review itself. While these resources acknowledge human-centered aspects, such as fairness and respectful communication, they treat them as secondary concern and do not explicitly address empathy, joint progress, and well-being of all peers involved. Moreover, they provide limited explicit, experience-based guidance on how reviewers can implement these values.

While this experience report aligns with the related work presented in reaffirming the importance of scientific care for quality, it differs by emphasizing the human-centered perspective of peer review --- advocating for collegial welfare for all peers involved. In this way, it contributes a complementary viewpoint to the state of the art, inviting reviewers to reconsider their role not merely as gatekeepers but as partners in the academic journey of each author, co-reviewer, workshop or conference organizer, and journal editor. This conceptual shift is essential for cultivating a constructive, ethically responsible, and empathetic peer review culture that advances our community.

\section{Conceptual Foundation for the Ethos and Guideline}
\label{sec:conceptual_foundation}
The conceptual foundation for the ethos of the PEERfect REVIEWer and its accompanying guideline with 16 practical recommendations is inspired by the widely known \textit{Extreme Programming} by Beck and Andres~\cite{Beck.2004} and the \textit{Golden Circle} by Sinek~\cite{Sinek.2009}.\vspace{0.1cm}

Inspired by the conceptual structure of \textit{Extreme Programming}~\cite{Beck.2004}, the ethos of the PEERfect \mbox{REVIEWer} is built on the relationship between \textit{values}, \textit{recommendations}, and \textit{actions}. \textit{Values} are abstract, universal criteria that individuals use to judge, decide, and justify their actions. Consequently, values set the direction for an individual’s actions by giving them meaning. \textit{Actions}, in contrast, are concrete, observable, and accountable --- they reflect how values are put into practice. However, values and actions are ``\textit{an ocean apart from each other}''~\cite[p. 14]{Beck.2004}. This gap is bridged by \textit{recommendations} which concretize values by providing guidance for an individual’s actions. Recommendations clarify what actions are intended to accomplish and help ensure that actions align with the underlying values.

To make the guidance even more accessible, I formulated the 16 practical recommendations using the pattern: ``[HOW] \textbf{to} [WHAT]. \textbf{Rationale:} [WHY].'' inspired by Sinek's \textit{Golden Circle}~\cite{Sinek.2009}. The \textit{Golden Circle} is a concept for guiding human behavior by helping others to understand ``\textit{why we do what we do}''~\cite[p. 42]{Sinek.2009}. In particular, Sinek~\cite{Sinek.2009} argues that this concept fosters so-called inspirational guidance, motivating individuals to take action, and is therefore a more powerful and sustainable way to influence human behavior. Accordingly, each recommendation explains how a PEERfect REVIEWer should proceed (HOW) to achieve a particular outcome (WHAT), while providing the rationale for following the recommendation (WHY).\vspace{0.1cm}

In this way, peer review practice is guided by the practical recommendations that help reviewers to take actions in alignment with the two core values of the ethos of the PEERfect REVIEWer.

\section{The Ethos of the PEERfect REVIEWer \& its Accompanying Guideline}
\label{sec:ethos_and_guideline}
Being a distinguished reviewer is often perceived as ``\textit{an innate talent or mystical art that requires decades of academic experience to master}''~\cite[p. 1]{Sridhar.2025}. The truth, however, is much simpler: Conduct peer review the way you expect it to be done. The PEERfect REVIEWer embraces this principle by embodying the two core values: \textit{Scientific care} and \textit{collegial welfare}. When a reviewer is committed to these core values, they conduct peer review with both scientific rigor and empathy, and thus (hopefully) as expected. While this ethos may seem abstract, it can be cultivated through guidance~\cite{Prechelt.2018, Sridhar.2025, Ernst.2021}.

\subsection{The PEERfect REVIEWer, its two Core Values, and Key Traits}
In the following, I first define the PEERfect REVIEWer and explain the two core values with their associated key traits to lay the foundation for the accompanying guideline with 16 practical recommendations for peer review with scientific rigor and empathy.

\vspace{-0.1cm}
\begin{mdframed}
     \textbf{Definition: The PEERfect REVIEWer} is a peer who exemplifies excellence in peer review by embodying \textit{scientific care} and \textit{collegial welfare}. They approaches peer review as a collaborative opportunity in which all peers involved work toward accepting publications rather than searching for reasons to reject them. They uphold high scientific standards and conduct peer review in a constructive, supportive, respectful, and timely manner. Through constructive engagement, ethical responsibility, and empathetic communication, they contribute meaningfully to the joint academic, professional, and personal progress and well-being of all peers involved --- including authors, co-reviewers, workshop or conference organizers, and journal editors.
\end{mdframed}
\vspace{-0.1cm}

\textit{Scientific care} reflects the reviewer's commitment as a caretaker for quality, upholding the quality and integrity of the submission under review. It encompasses critical assessment of each submission with scientific rigor regarding novelty, soundness \& rigor, relevance \& impact, verifiability \& transparency, presentation, community value \& perspective (cf. \autoref{app:common_criteria}), and its compliance with the standards of the discipline. Scientific care is expressed through the following key traits: Attention to detail, constructiveness, fairness, objectivity, and factual accuracy. A reviewer practicing scientific care engages deeply with each submission and provides detailed, constructive, fair, proportionate, and well-justified feedback, ensuring that criticism is grounded in reason rather than opinion. This value safeguards the credibility and trust in the peer review process and supports joint academic and professional progress through rigorous and responsible engagement with each submission.

\textit{Collegial welfare} reflects the reviewer's commitment as a steward of joint progress and well-being, upholding empathy and respect throughout the entire peer review process. It encompasses thoughtful engagement with all peers involved through kind and supportive communication, reliable interaction, and a genuine intent to help others. Collegial welfare is expressed through the following key traits: Humility, curiosity, reliability, kindness, and helpfulness. A reviewer practicing collegial welfare acknowledges the limits of their expertise, avoids overstepping their bounds, and engages with each submission in a spirit of inquiry rather than judgment. They offer timely feedback in a kind and encouraging tone, along with actionable suggestions. This value contributes to recognizing the human effort behind each submission and remaining mindful of the emotional and professional impact of words, ensuring psychological safety, mutual respect, and joint professional and personal progress.

\subsection{An Accompanying Guideline for Peer Review with Scientific Rigor and Empathy}
Below, I present the accompanying guideline with the 16 practical recommendations that help reviewers align their actions with the two core values for peer review: \textit{Scientific care} and \textit{collegial welfare}. Each recommendation follows the pattern: ``[HOW] \textbf{to} [WHAT]. \textbf{Rationale:} [WHY].''. Beyond improving review quality, the recommendations are intended to foster empathy, joint progress, and well‑being of all peers involved. The rationales make these human‑centered values explicit by explaining how each recommendation contributes to a constructive, respectful, and collaborative peer review culture. To support implementation, \autoref{app:examples} provides positive and negative examples for each recommendation. For clarity and ease of use, I also grouped the recommendations into four themes: \textit{Commitment \& Planning}, \textit{Review Preparation \& Execution}, \textit{Feedback \& Tone}, and \textit{Ethical Responsibility \& Boundaries}.\vspace{0.1cm}

\noindent
\textit{Commitment \& Planning}: Establishing a foundation for timely and responsible peer review practice.
\begin{enumerate}[noitemsep, topsep=0pt]
    \item Accept a review invitation only if you have sufficient time \textbf{to} ensure a timely peer review process.
    
    \textbf{Rationale:} Saying ``no'' early helps all peers involved more than receiving no, delayed, or poor reviews, which create stress, disrupt planning, and negatively affect their well‑being.
    
    \item Track all relevant deadlines \textbf{to} plan your review workload effectively.
    
    \textbf{Rationale:} Awareness of submission, bidding, review, discussion, rebuttal, and meta‑review deadlines enables responsible time allocation, reduces last‑minute pressure, and supports a predictable and considerate process for all peers involved.
    
    \item Schedule dedicated review sessions and communicate proactively with workshop or conference organizers and journal editors if issues arise \textbf{to} maintain timeliness and transparency.
    
    \textbf{Rationale:} Timely and transparent communication demonstrates respect, reduces uncertainty, and supports the emotional well‑being of all peers involved by preventing avoidable delays and misunderstandings.\vspace{0.1cm}
\end{enumerate}

\noindent
\textit{Review Preparation \& Execution}: Ensuring effective reviewing.
\begin{enumerate}[noitemsep, topsep=0pt]
    \setcounter{enumi}{3}
    \item Use a structured review template \textbf{to} organize your feedback clearly and effectively.
    
    \textbf{Rationale:} A consistent format (e.g., summary, strengths, weaknesses, detailed comments, and minor comments) presents your feedback clearly and coherently, improving readability, supporting decision‑making, and reducing cognitive load for all peers involved.
    
    \item Integrate the review criteria provided into your review template \textbf{to} ensure your review aligns with the workshop, conference, or journal expectations.
    
    \textbf{Rationale:} Integrating the review criteria ensures fairness, consistency, and clarity, enabling all peers involved to understand your review more easily. Clear alignment with the criteria reduces cognitive load and supports a smoother, more collaborative peer review process.
    
    \item Allocate 1–12 hours to read, understand, and assess the submission \textbf{to} provide detailed and constructive feedback.
    
    \textbf{Rationale:} Deep engagement with the submission ensures scientific rigor and demonstrates respect for the authors’ work, fostering joint progress and reducing the emotional stress caused by superficial or uninformed reviews. The time required naturally varies with the length and complexity of the submission, so allocating sufficient time supports all peers involved by providing a reliable basis for fair and efficient decision‑making.

    \item Evaluate any provided artifact \textbf{to} assess the submission's replicability and transparency.
    
    \textbf{Rationale:} Artifact evaluation strengthens scientific integrity and supports authors’ efforts toward open science, while also enabling co-reviewers, workshop or conference organizers, and journal editors to make fair and well‑informed decisions.\vspace{0.1cm}
\end{enumerate}

\noindent
\textit{Feedback \& Tone}: Delivering constructive, respectful, and actionable feedback.
\begin{enumerate}[noitemsep, topsep=0pt]
    \setcounter{enumi}{7}
    \item Justify your feedback with clear references to the submission and, when relevant, to the literature \textbf{to} enhance credibility and trust.
    
    \textbf{Rationale:} Traceable and well‑justified feedback helps all peers involved understand your perspective more easily, reduces misunderstandings, and supports an efficient peer review process.
    
    \item Use respectful, unbiased, and non-dismissive language \textbf{to} foster psychological safety.
    
    \textbf{Rationale:} Respectful communication acknowledges the effort behind every submission, reduces emotional stress, and encourages all peers involved to engage openly and constructively with your feedback, enabling a professional and collaborative dialogue throughout the review process.
    
    \item Acknowledge the strengths of the submission \textbf{to} highlight its contributions.
    
    \mbox{\textbf{Rationale:}} Highlighting strengths ensures that the contributions of a submission are visible and not overshadowed by weaknesses. In this way, authors feel recognized for their efforts, and co‑reviewers, workshop or conference organizers, and journal editors gain a clear understanding of the submission’s contributions, enabling them to make balanced and well‑informed decisions.
    
    \item Offer actionable suggestions using the WHAT–WHY–HOW approach \textbf{to} promote clarity, improvement, and usefulness.
    
    \textbf{Rationale:} Actionable suggestions require stating WHAT you observed, WHY the observation matters for the submission’s quality or validity, and HOW the authors could address it in a constructive and achievable way. In this way, you reduce ambiguity and support authors’ understanding by making improvement feel achievable. Co-reviewers, workshop or conference organizers, and journal editors can also more easily understand how identified issues relate to the submission’s quality and how they could be addressed.\vspace{0.1cm}
\end{enumerate}

\noindent
\textit{Ethical Responsibility \& Boundaries}: Ensuring integrity and accountability in the peer review process.
\begin{enumerate}[noitemsep, topsep=0pt]
    \setcounter{enumi}{11}
    \item Accept a review invitation only if you possess domain-specific expertise \textbf{to} ensure a qualified and credible assessment. 
    
    \textbf{Rationale:} Reviewing outside your expertise carries the risk of an incorrect assessment of the submission and undermines fairness, which can harm authors’ well‑being and erode trust among all peers involved.
    
    \item Declare any conflicts of interest and withdraw if necessary \textbf{to} preserve impartiality and trust.
    
    \textbf{Rationale:} Transparency about potential conflicts of interest protects the integrity and fairness of the review process by safeguarding all peers involved from biased or compromised assessments.

    \item Reflect on potential researcher biases \textbf{to} ensure a fair and balanced assessment.
    
    \textbf{Rationale:} Personal, methodological, and disciplinary preferences can unintentionally influence your assessment. Reflecting on these potential biases helps ensure that your assessment focuses on the submission’s actual contributions rather than on your own preferences. In this way, all peers involved can be confident that your review is fair, balanced, and grounded in the submission~itself.

    \item Disclose if you involve colleagues as sub-reviewers \textbf{to} uphold accountability for the review. 
    
    \mbox{\textbf{Rationale:}} Sub‑reviewing fosters learning and mentoring, but you remain fully accountable for the review and must disclose any sub‑reviewers to maintain trust among all peers involved.

    \item Refrain from using generative artificial intelligence (AI) to produce reviews \textbf{to} ensure human assessment and respect intellectual boundaries.
    
    \textbf{Rationale:} Publishers generally prohibit AI‑generated reviews, and sending a submission to an AI system violates the authors’ intellectual property and the confidentiality of their unpublished work. By accepting the review, you have consciously agreed to read, understand, and assess the submission yourself, and delegating this responsibility to an AI compromises the integrity of the review for all peers involved, who rely on your independent and accountable assessment.
\end{enumerate}

By following this guideline, reviewers contribute to a constructive, ethically responsible, and empathetic peer review culture --- acting as PEERfect REVIEWers with scientific care and collegial welfare.

\section{Discussion}
\label{sec:discussion}
This experience report enriches the ongoing discourse on peer review in SE and initiates this discussion within the RE community by presenting the ethos of the PEERfect REVIEWer and its accompanying guideline for peer review with scientific rigor and empathy. They offer explicit, experience-based guidance to promote impactful peer review practice that upholds high scientific standards in a constructive, supportive, respectful, and timely manner. Given their generic formulation, the ethos and guideline are broadly applicable --- supporting both early-career researchers seeking orientation and experienced reviewers reflecting on their practice. By balancing scientific care and collegial welfare, they reframe peer review as a collaborative opportunity in which all peers involved work toward accepting publications rather than searching for reasons to reject them. In this way, authors, reviewers, workshop or conference organizers, and journal editors together ensure high-quality publications and foster joint progress and well-being.

This experience report responds to a growing need within our community for explicit, experienced-based guidance. While existing literature primarily emphasizes the technical dimensions of peer review --- often framing it as a gatekeeping mechanism --- peer-reviewed literature that contributes a human-centered perspective remains limited. This report addresses this gap by positioning empathy as a complementary force to scientific rigor in cultivating a constructive, ethically responsible, and empathetic peer review culture. In this way, it advances our community toward a more holistic understanding of peer review as both a scientific and human endeavor.

The ethos of the PEERfect REVIEWer and its accompanying guideline are designed to be broadly applicable across venues and communities. Nevertheless, they may not fully capture the nuances of all disciplinary contexts, cultural expectations, or reviewing traditions. Their successful adoption depends on individual motivation and institutional support. Reviewers under time pressure may struggle to implement all recommendations, especially those requiring deeper engagement and reflection. However, the guideline aims to mitigate time pressure through proactive planning and responsible commitment --- helping reviewers manage their workload before it becomes overwhelming. Without incentives, such as editorial recognition or distinguished reviewer awards, and integration into reviewer onboarding and training, uptake may remain limited. The ethos and guideline promote impactful peer review practice, but they cannot enforce it; their success and impact rely on voluntary engagement and a broader conceptual shift within our community. I build on the preliminary guideline by Ernst et al.~\cite{Ernst.2021}, which was ``\textit{naturally far from being perfect, let alone complete}''~\cite[p. 27]{Ernst.2021} to offer an extension toward a more comprehensive and human-centered perspective of peer review. While informed by literature and practice, this experience report reflects my personal perspective within our community. Further application, adaptation, and collective refinement are needed to move closer to achieving true ``PEERfection'' and completion.

In this regard, short-term actions should focus on active dissemination to raise awareness and encourage adoption. First results have already been achieved: After a \href{https://www.linkedin.com/posts/oliver-karras_ready-to-become-the-%F0%9D%97%A3%F0%9D%97%98%F0%9D%97%98%F0%9D%97%A5fect-%F0%9D%97%A5%F0%9D%97%98-activity-7353768250448486400-oQ_v}{\textit{\textcolor{link}{talk on ``The PEERfect REVIEWer''}}}, \href{https://conf.researchr.org/profile/seams-2026/amelbennaceur}{\textit{\textcolor{link}{Amel Bennaceur, Program Committee co-Chair of SEAMS 2026}}}, requested to include the guideline in her slides for the program committee’s kick-off meeting. Mid-term actions should embed the ethos and guideline into onboarding and training across workshops, conferences, and journals, and evaluate their impact through empirical studies. The REFSQ conference, with its explicit commitment to supportive, inclusive, and high‑quality reviewing\footnote{Excerpt from the invitation email to the REFSQ 2026 program committee: ``\textit{Please note that the goal is to foster the REFSQ community by being as supportive and inclusive as possible in the reviewing process. We collectively need to make an effort to provide detailed and high-quality reviews, giving our reviews the same consideration and care that we would want to receive from reviewers of our own submissions.}''} offers a promising venue for piloting and refining the ethos and guideline. Long-term actions should establish a community-maintained version of the guideline, similar to other established guidelines (cf. Section~\ref{sec:rw}), and advocate for institutional recognition of remarkable peer review practice with scientific rigor and empathy. Together, these actions help cultivate a peer review culture that advances our community by being grounded in scientific care and collegial welfare.

\section{Conclusion}
\label{sec:conclusion}
Peer review cannot be reduced to its technical dimensions alone; at its core, it is a human endeavor --- a collaborative opportunity to shape the quality of scientific work and foster joint progress and well-being of all peers involved. The ethos of PEERfect REVIEWer and its accompanying guideline offer a starting point for cultivating a peer review culture that balances scientific care and collegial welfare. Like software development, peer review is a complex socio-technical system involving interdependent stakeholders. Just as software cannot succeed without engaging the people it serves, peer review cannot thrive without acknowledging its human dimension. Reviewers must reconsider their role not merely as gatekeepers, but as partners in the academic journey of each author, co-reviewer, workshop or conference organizer, and journal editor. As Prechelt et al.~\cite[pp. 84-85]{Prechelt.2018} aptly noted: ``\textit{Time investment is a matter of priorities, lack of expertise can be accommodated by not taking on the review in the first place, and a lack of} [scientific] \textit{care} [and collegial welfare] \textit{stems from a modifiable (if not easily) attitude. Improvement efforts can be successful in principle}'' --- if we decide to make them. To advance our community through the pursuit of ``PEERfection'', we must not only ``embrace change'', but also embrace our peers --- recognizing scientific rigor and empathy as complementary forces that promote impactful peer review practice.


\section*{Declaration on Generative AI}
The author used Microsoft Copilot and Grammarly for grammar and spelling check. The author reviewed and edited the content as needed and take full responsibility for the publication’s content. 

\bibliography{sample-ceur}

\appendix

\section{Common Review Criteria in SE and RE}\label{app:common_criteria}
This appendix provides a distilled set of unified review criteria used across established SE and RE conferences, including \href{https://conf.researchr.org/track/icse-2026/icse-2026-research-track#review-criteria}{\textit{\textcolor{link}{ICSE}}}, \href{https://conf.researchr.org/track/ase-2026/ase-2026-research-track#review-criteria}{\textit{\textcolor{link}{ASE}}}, \href{https://conf.researchr.org/track/RE-2026/RE-2026-Research-Papers#review-criteria}{\textit{\textcolor{link}{RE}}}, and \href{https://2026.refsq.org/track/refsq-2026-research-papers#Review-Criteria}{\textit{\textcolor{link}{REFSQ}}}. (see \autoref{tab:unified_criteria}). For each unified criterion, I offer a consolidated definition with guiding questions for its assessment. In this way, reviewers receive a checklist and a shared understanding of what typically needs to be assessed in SE and RE, generally independent of specific paper types or conference traditions. However, some criteria include optional guiding questions tailored to specific paper types, namely vision papers and research previews. These questions are clearly marked and do not affect the general applicability of the unified review criteria.\vspace{0.1cm}

\noindent
\textbf{1. Novelty:}\\
The extent to which the problem formulation, methodology, proposed solution, or evaluation introduces ideas that are innovative and meaningfully distinct from prior work. This criterion includes originality relative to state‑of‑the‑art research and literature.

\newpage

\textbf{Guiding questions:}
\begin{itemize}[noitemsep, topsep=0pt]
    \item To what extent is the proposed solution, study, or vision novel with respect to the state of the art and existing literature?
    \item Did the authors clearly identify related work and articulate their unique contribution?
    \item Does the study address a contemporary problem and extend the existing body of knowledge?
    \item \textit{For vision papers / research previews}: Does the paper make you think, ``\textit{I heard it first at this conference}''?
\end{itemize}
\vspace{0.2cm}

\noindent
\textbf{2. Soundness \& Rigor:}
The technical depth, methodological appropriateness, and logical validity of the work. This criterion includes whether claims and conclusions are supported by appropriate research methods, justified methodological choices, rigorous data analysis, and a transparent discussion of limitations and threats to validity.\vspace{0.1cm}

\textbf{Guiding questions:}
\begin{itemize}[noitemsep, topsep=0pt]
    \item Has the proposed solution been developed using a well‑motivated approach and recognized research methods?
    \item Are the methodological and design choices justified and appropriate for the problem at hand?
    \item Did the authors clearly state their research questions, data collection procedures, and data analysis methods?
    \item Are the claims and conclusions logically derived from the data and supported by rigorous evidence?
    \item Did the authors transparently discuss limitations and threats to validity?
    \item \textit{For vision papers / research previews}: Is the vision / research idea supported by coherent arguments, or did the authors provide a convincing proof‑of‑concept?
\end{itemize}

\begin{table}[t]
\centering
\caption{Unified review criteria synthesized from the official review guidelines of ICSE, ASE, RE, and REFSQ.\protect\footnotemark}
\label{tab:unified_criteria}
\resizebox{\textwidth}{!}{%
\begin{tabular}{|l||llll|}
\hline
\multicolumn{1}{|c||}{\textbf{Unified Criterion}} & \multicolumn{1}{c|}{\textbf{ICSE}} & \multicolumn{1}{c|}{\textbf{ASE}} & \multicolumn{1}{c|}{\textbf{RE}} & \multicolumn{1}{c|}{\textbf{REFSQ}} \\ \hline \hline
\textbf{Novelty} & \multicolumn{1}{l|}{Novelty} & \multicolumn{1}{l|}{Novelty} & \multicolumn{1}{l|}{Novelty} & Novelty \\ \hline
\textbf{\begin{tabular}[c]{@{}l@{}}Soundness \&\\ Rigor\end{tabular}} & \multicolumn{1}{l|}{Rigor} & \multicolumn{1}{l|}{\begin{tabular}[c]{@{}l@{}}Soundness / \\ Insights and Evidence\end{tabular}} & \multicolumn{1}{l|}{Soundness} & \begin{tabular}[c]{@{}l@{}}Soundness / \\ Soundness of Arguments /\\ Soundness of the Research Plan\end{tabular} \\ \hline
\textbf{\begin{tabular}[c]{@{}l@{}}Relevance \&\\ Impact\end{tabular}} & \multicolumn{1}{l|}{Relevance} & \multicolumn{1}{l|}{\begin{tabular}[c]{@{}l@{}}Significance / \\ Importance and Scope\end{tabular}} & \multicolumn{1}{l|}{Potential Impact} & \begin{tabular}[c]{@{}l@{}}Potential Impact /\\ Relevance /\\ Relevance of the Application\end{tabular} \\ \hline
\textbf{\begin{tabular}[c]{@{}l@{}}Verifiability \&\\ Transparency\end{tabular}} & \multicolumn{1}{l|}{\begin{tabular}[c]{@{}l@{}}Verifiability and\\ Transparency\end{tabular}} & \multicolumn{1}{l|}{Verifiability} & \multicolumn{1}{l|}{Verifiability} & Verifiability \\ \hline
\textbf{Presentation} & \multicolumn{1}{l|}{Presentation} & \multicolumn{1}{l|}{Presentation} & \multicolumn{1}{l|}{Presentation} & Presentation \\ \hline
\textbf{\begin{tabular}[c]{@{}l@{}}Community Value \&\\ Perspective\end{tabular}} & \multicolumn{1}{l|}{\textit{N/A}} & \multicolumn{1}{l|}{Perspective} & \multicolumn{1}{l|}{\textit{N/A}} & Potential for Discussion \\ \hline
\end{tabular}%
}
\vspace{-0.2cm}
\end{table}

\footnotetext{The \href{https://conf.researchr.org/home/fse-2026}{\textit{\textcolor{link}{ACM International Conference on the Foundations of Software Engineering}}} (FSE) is one of the top-3 leading SE conferences, but it does not publicly provide official review criteria; therefore, it is not included in this comparison.}

\noindent
\textbf{3. Relevance \& Impact:}\\
The significance of the research to the discipline. This criterion includes the importance of the addressed problem, the potential influence on future research and practice, and the clarity with which the contribution’s value is articulated.\vspace{0.1cm}

\textbf{Guiding questions:}
\begin{itemize}[noitemsep, topsep=0pt]
    \item Is the significance of the research clearly stated, and is its potential impact on research and practice convincing?
    \item Does the paper address a problem of practical importance, and is the application context sufficiently representative?
    \item Has the proposed solution been evaluated --- even preliminarily --- in a realistic or representative setting?
    \item \textit{For vision papers / research previews}: Do the authors outline a roadmap and articulate both short‑term and long‑term impacts of their vision / research idea?
\end{itemize}

\noindent
\textbf{4. Verifiability \& Transparency:}\\
The degree to which the work enables independent verification, reuse, or replication of its results. This criterion includes the availability and clarity of data, software, artifacts, and methodological details that support the credibility and reproducibility of the work.\vspace{0.1cm}

\textbf{Guiding questions:}
\begin{itemize}[noitemsep, topsep=0pt]
    \item Does the paper provide sufficient information to understand how the innovation works and how the evidence or data was produced?
    \item Did the authors share software, data, and supplementary materials?
    \item Are there clear guidelines for reusing artifacts and replicating the results?
    \item Overall, does the paper adequately support independent verification and replication of its contributions?
\end{itemize}
\vspace{0.1cm}

\noindent
\textbf{5. Presentation:}\\
The quality of the work's writing, structure, and overall communication. This criterion includes clarity of problem statements, methodological descriptions, and results; effective use of readable figures and tables; adequate language usage; and the accessibility of technical content for the intended audience.\vspace{0.1cm}

\textbf{Guiding questions:}
\begin{itemize}[noitemsep, topsep=0pt]
    \item Is the paper clearly presented, well‑structured, and free of major ambiguity?
    \item Is the writing of high quality, including adequate language usage, and readable figures and tables?
    \item To what extent can the content be understood by the broader community?
    \item If highly technical content is presented, did the authors make a deliberate effort to summarize their proposal or study in an intuitive way?
\end{itemize}
\vspace{0.2cm}

\noindent
\textbf{6. Community Value \& Perspective:}\\
The extent to which the work offers actionable insights, stimulates discussion, or provides forward‑looking perspectives that can inspire researchers and practitioners. This criterion includes the articulation of lessons learned, contextual reflections, and thought‑provoking ideas that enrich the community’s understanding.\vspace{0.1cm}

\textbf{Guiding questions:}
\begin{itemize}[noitemsep, topsep=0pt]
    \item Are the lessons learned sufficiently insightful, clearly described, and supported by evidence (including anecdotal evidence where appropriate)?
    \item Will the presentation of this paper raise audience interest and spark meaningful discussion?
    \item Could the paper provoke constructive controversy or elicit deeper reflection within the community?
    \item To what extent can conference participants draw inspiration from this work to develop novel solutions or conduct sound empirical evaluations of their own?
\end{itemize}

\section{Positive and Negative Examples for Each Recommendation}\label{app:examples}
Translating the 16 practical recommendations of the accompanying guideline for peer review with scientific rigor and empathy into concrete reviewing behavior can be challenging. This appendix provides short, practice‑oriented examples to support reviewers in applying the recommendations. For each recommendation, I show a positive instance of good reviewing and a negative one that illustrates common pitfalls. These examples help reviewers calibrate their tone, avoid unintended dismissiveness, and contribute to a constructive, ethically responsible, and empathetic peer review culture.\vspace{0.1cm}

\noindent
\textit{Commitment \& Planning}:
\begin{enumerate}[noitemsep, topsep=0pt]
    \item Accept a review invitation only if you have sufficient time \textbf{to} ensure a timely peer review process.
    
    \textbf{Rationale:} Saying ``no'' early helps all peers involved more than receiving no, delayed, or poor reviews, which create stress, disrupt planning, and negatively affect their well‑being.\vspace{0.1cm} 
    
    \textbf{Positive example}: Replying within 48 hours: ``Thank you for the invitation. Unfortunately, my current workload prevents me from committing the necessary time to provide a thorough review by the deadline. I must decline.''
    
    \textbf{Negative example:} Accepting the invitation with good intentions, realizing a week before the deadline you have no time, and ultimately ignoring the deadline or submitting a short review.\vspace{0.1cm}

   \item Track all relevant deadlines \textbf{to} plan your review workload effectively.
   
   \textbf{Rationale:} Awareness of submission, bidding, review, discussion, rebuttal, and meta‑review deadlines enables responsible time allocation, reduces last‑minute pressure, and supports a predictable and considerate process for all peers involved.\vspace{0.1cm}
    
   \textbf{Positive example:} Upon accepting a review, immediately blocking out calendar time for reading the paper, writing the review, and participating in the rebuttal/discussion phases.
   
   \textbf{Negative example:} Forgetting about the review until the automated reminder email arrives 24 hours before the deadline, forcing a rushed and stressful assessment.\vspace{0.1cm}

   \item Schedule dedicated review sessions and communicate proactively with workshop or conference organizers and journal editors if issues arise \textbf{to} maintain timeliness and transparency.
    
   \textbf{Rationale:} Timely and transparent communication demonstrates respect, reduces uncertainty, and supports the emotional well‑being of all peers involved by preventing avoidable delays and misunderstandings.\vspace{0.1cm}
    
   \textbf{Positive example:} Emailing the track chair two weeks before the deadline: ``Due to an unexpected medical emergency, I will need an extension of three days. Let me know if this works or if you need to reassign the paper.''
   
   \textbf{Negative example:} Ghosting the chairs, missing the deadline entirely, and only responding after the chairs have frantically sent multiple follow-up emails.\vspace{0.1cm}
\end{enumerate}

\noindent
\textit{Review Preparation \& Execution}:
\begin{enumerate}[noitemsep, topsep=0pt]
    \setcounter{enumi}{3}
      \item Use a structured review template \textbf{to} organize your feedback clearly and effectively.
    
    \textbf{Rationale:} A consistent format (e.g., summary, strengths, weaknesses, detailed comments, and minor comments) presents your feedback clearly and coherently, improving readability, supporting decision‑making, and reducing cognitive load for all peers involved.\vspace{0.1cm}
    
    \textbf{Positive example:} Organizing the review with clear headings, e.g., Summary, Strengths, Weaknesses, Detailed Comments, Minor Comments.
    
    \textbf{Negative example:} Submitting a single, massive block of text where fundamental methodological critiques are randomly mixed with complaints about missing commas.\vspace{0.1cm}

    \item Integrate the review criteria provided into your review template \textbf{to} ensure your review aligns with the workshop, conference, or journal expectations.
    
    \textbf{Rationale:} Integrating the review criteria ensures fairness, consistency, and clarity, enabling all peers involved to understand your review more easily. Clear alignment with the criteria reduces cognitive load and supports a smoother, more collaborative peer review process.\vspace{0.1cm}
    
    \textbf{Positive example:} Explicitly linking feedback to the venue's review criteria: ``Regarding the `Novelty' criterion, the submission clearly demonstrates...''
    
    \textbf{Negative example:} Evaluating a short, visionary workshop paper based on the rigorous empirical standards expected of a 20-page journal article.\vspace{0.1cm}

    \item Allocate 1–12 hours to read, understand, and assess the submission \textbf{to} provide detailed and constructive feedback.
    
    \textbf{Rationale:} Deep engagement with the submission ensures scientific rigor and demonstrates respect for the authors’ work, fostering joint progress and reducing the emotional stress caused by superficial or uninformed reviews. The time required naturally varies with the length and complexity of the submission, so allocating sufficient time supports all peers involved by providing a reliable basis for fair and efficient decision‑making.\vspace{0.1cm}
    
    \textbf{Positive example}: Taking the time to read the paper twice, check a few key references, and write a thoughtful review.

    \textbf{Negative example:} Skimming only the abstract and conclusion 30 minutes before the deadline and submitting a generic ``Good paper, accept'' review.\vspace{0.1cm}

    \item Evaluate any provided artifact \textbf{to} assess the submission's replicability and transparency.
    
    \textbf{Rationale:} Artifact evaluation strengthens scientific integrity and supports authors’ efforts toward open science, while also enabling co-reviewers, workshop or conference organizers, and journal editors to make fair and well‑informed decisions.\vspace{0.1cm}
    
    \textbf{Positive example:} Navigating to the provided GitHub repository, checking if the ReadMe instructions are clear, and verifying that the dataset is actually accessible.

    \textbf{Negative example:} Complaining in the review that ``the authors did not explain how they filtered the data'', when the authors provided a fully documented script in their open science repository that the reviewer ignored.\vspace{0.1cm}
\end{enumerate}

\noindent
\textit{Feedback \& Tone}:
\begin{enumerate}[noitemsep, topsep=0pt]
    \setcounter{enumi}{7}
     \item Justify your feedback with clear references to the submission and, when relevant, to the literature \textbf{to} enhance credibility and trust.
    
    \textbf{Rationale:} Traceable and well‑justified feedback helps all peers involved understand your perspective more easily, reduces misunderstandings, and supports supports an efficient peer review process.\vspace{0.1cm}
    
    \textbf{Positive example:} ``The conclusion on page 8 states that `Factor X is the primary driver of success'. However, the data presented in Table 4 (page 5) appears to show that Factor Y has a much higher statistical significance. Clarify this discrepancy or provide the underlying analysis that supports the focus on Factor X.''
    
    \textbf{Negative example:} ``The authors' conclusions are not supported by their data and the logic in Section 5 is flawed.'' (Without pointing to the specific contradiction in the table or step).\vspace{0.1cm}
    
    \item Use respectful, unbiased, and non-dismissive language \textbf{to} foster psychological safety.
    
    \textbf{Rationale:} Respectful communication acknowledges the effort behind every submission, reduces emotional stress, and encourages all peers involved to engage openly and constructively with your feedback, enabling a professional and collaborative dialogue throughout the review process.\vspace{0.1cm}
    
    \textbf{Positive example:} ``The methodology could be strengthened by clarifying the participant selection criteria, as the current description leaves it open to selection bias.''

    \textbf{Negative example:} ``The authors clearly have no idea how to conduct empirical research. The methodology is a complete mess.''\vspace{0.1cm}

    \item Acknowledge the strengths of the submission \textbf{to} highlight its contributions.
    
    \mbox{\textbf{Rationale:}} Highlighting strengths ensures that the contributions of a submission are visible and not overshadowed by weaknesses. In this way, authors feel recognized for their efforts, and co‑reviewers, workshop or conference organizers, and journal editors gain a clear understanding of the submission’s contributions, enabling them to make balanced and well‑informed decisions.\vspace{0.1cm}
    
    \textbf{Positive example}: Starting the review with: ``The authors tackle a highly relevant problem in the community, and the dataset collected represents a valuable contribution to the field.''

    \textbf{Negative example:} Skipping the summary entirely and starting the review with: ``Weaknesses: The writing is terrible and the graphs are unreadable.''\vspace{0.1cm}

    \item Offer actionable suggestions using the WHAT–WHY–HOW approach \textbf{to} promote clarity, improvement, and usefulness.
    
    \textbf{Rationale:} Actionable suggestions require stating WHAT you observed, WHY the observation matters for the submission’s quality or validity, and HOW the authors could address it in a constructive and achievable way. In this way, you reduce ambiguity and support authors’ understanding by making improvement feel achievable. Co-reviewers, workshop or conference organizers, and journal editors can also more easily understand how identified issues relate to the submission’s quality and how they could be addressed.\vspace{0.1cm}
    
    \textbf{Positive example:} ``Section 4 lacks a discussion on threats to construct validity (WHAT). This makes it difficult to assess if the survey questions actually measured what was intended (WHY). Add a paragraph detailing how the survey instrument was validated prior to distribution (HOW).''

    \textbf{Negative: example:} ``Add a threats to validity section.''\vspace{0.1cm}
\end{enumerate}

\noindent
\textit{Ethical Responsibility \& Boundaries}:
\begin{enumerate}[noitemsep, topsep=0pt]
    \setcounter{enumi}{11}
    \item Accept a review invitation only if you possess domain-specific expertise \textbf{to} ensure a qualified and credible assessment. 
    
    \textbf{Rationale:} Reviewing outside your expertise carries the risk of an incorrect assessment of the submission and undermines fairness, which can harm authors’ well‑being and erode trust among all peers involved.\vspace{0.1cm}
    
    \textbf{Positive example:} Declining to review a paper that heavily focuses on formal methods and theorem proving because your expertise lies strictly in qualitative, human-centric methodologies.

    \textbf{Negative example:} Accepting a paper completely outside your wheelhouse just to ``learn about the topic'', resulting in a superficial review that misses critical technical flaws.\vspace{0.1cm}

    \item Declare any conflicts of interest and withdraw if necessary \textbf{to} preserve impartiality and trust.
    
    \textbf{Rationale:} Transparency about potential conflicts of interest protects the integrity and fairness of the review process by safeguarding all peers involved from biased or compromised assessments.\vspace{0.1cm}
    
    \textbf{Positive example:} Realizing during the bidding phase that a paper is authored by a former PhD student you published with two years ago, and explicitly declaring a conflict in the review system.

    \textbf{Negative example:} Reviewing a close collaborator's paper and giving it a ``Strong Accept'' to boost their publication record.\vspace{0.1cm}

    \item Reflect on potential researcher biases \textbf{to} ensure a fair and balanced assessment.
    
    \textbf{Rationale:} Personal, methodological, and disciplinary preferences can unintentionally influence your assessment. Reflecting on these potential biases helps ensure that your assessment focuses on the submission’s actual contributions rather than on your own preferences. In this way, all peers involved can be confident that your review is fair, balanced, and grounded in the submission~itself.\vspace{0.1cm}
    
    \textbf{Positive example:} Recognizing that while you personally prefer grounded theory, the authors' choice of a case study approach is valid, well-executed, and appropriate for their research question.

    \textbf{Negative example:} Recommending a rejection simply because the authors used qualitative methods, and you personally believe only quantitative, statistical methods hold scientific value.\vspace{0.1cm}

    \item Disclose if you involve colleagues as sub-reviewers \textbf{to} uphold accountability for the review. 
    
    \mbox{\textbf{Rationale:}} Sub‑reviewing fosters learning and mentoring, but you remain fully accountable for the review and must disclose any sub‑reviewers to maintain trust among all peers involved.\vspace{0.1cm}
    
    \textbf{Positive example:} Including a confidential note to the chairs: ``Please note that this review was co-written with my PhD student, as part of her peer-review training.''
    
    \textbf{Negative example:} Handing the paper to a junior PhD student, letting them write the entire review, and submitting it under your own name without giving them credit or notifying the chairs.\vspace{0.1cm}

    \item Refrain from using generative artificial intelligence (AI) to produce reviews \textbf{to} ensure human assessment and respect intellectual boundaries.
    
    \textbf{Rationale:} Publishers generally prohibit AI‑generated reviews, and sending a submission to an AI system violates the authors’ intellectual property and the confidentiality of their unpublished work. By accepting the review, you have consciously agreed to read, understand, and assess the submission yourself, and delegating this responsibility to an AI compromises the integrity of the review for all peers involved, who rely on your independent and accountable assessment.\vspace{0.1cm}
    
    \textbf{Positive example:} Reading the submission yourself, synthesizing your own thoughts, and writing the review based on your human expertise and understanding of the field's nuances.

    \textbf{Negative example:} Uploading the unpublished, confidential PDF to ChatGPT, asking it to ``write a peer review for this paper'', and copy-pasting the output into the review form.
\end{enumerate}

\end{document}